\def\umass{1}
\def\austin{2}

\documentclass[iop]{emulateapj}
\usepackage{natbib}
\usepackage{graphicx}
\usepackage[space]{grffile}
\usepackage{latexsym}
\usepackage{amsfonts,amsmath,amssymb}
\usepackage{url}
\usepackage[utf8]{inputenc}
\usepackage{fancyref}
\usepackage[colorlinks=true,linkcolor=blue,citecolor=blue]{hyperref}

\usepackage{ulem} %RTF: for strike-through in resubmission
\usepackage{color} %RTF : For color in resubmission

\begin{document}

\title{Single-Degenerate Type Ia Supernovae Are Preferentially Overluminous}

%% Notice placement of commas and superscripts and use of &
%% in the author list

\author{Robert Fisher$^{1}$ \& Kevin Jumper$^{2}$}

%\maketitle

\altaffiltext {\umass} {Department of Physics, University of Massachusetts Dartmouth, 285 Old Westport Road, North Dartmouth, Ma. 02740, USA}
\altaffiltext {\austin} {Department of Astronomy, University of Texas at Austin,  2515 Speedway, Austin, Tx., 78712-1205, USA}

\begin {abstract}

%%Rates, delay times, AND host environments

Recent observational and theoretical progress has favored merging and helium-accreting sub-Chandrasekhar mass white dwarfs in the double-degenerate and the double-detonation channels, respectively, as the most promising progenitors of normal Type Ia supernovae (SNe Ia). Thus the fate of rapidly-accreting Chandrasekhar mass white dwarfs in the single-degenerate channel remains more mysterious then ever. In this paper, we clarify the nature of ignition in Chandrasekhar-mass single-degenerate SNe Ia by analytically deriving the existence of a characteristic length scale which establishes a transition from central ignitions to buoyancy-driven ignitions. Using this criterion, combined with data from three-dimensional simulations of convection and ignition, we demonstrate that the overwhelming majority of ignition events within Chandrasekhar-mass white dwarfs in the single-degenerate channel are buoyancy-driven, and consequently lack a vigorous deflagration phase.   We thus infer that single-degenerate SNe Ia are generally expected to lead to overluminous 1991T-like SNe Ia events. We establish that the rates predicted from both the population of supersoft X-ray sources and binary population synthesis models of the single-degenerate channel are broadly consistent with the observed rates of overluminous SNe Ia, and suggest that the population of supersoft X-ray sources are the dominant stellar progenitors of SNe 1991T-like events. We further demonstrate that the single-degenerate channel contribution to the normal and failed 2002cx-like rates is not likely to exceed 1\% of the total SNe Ia rate. We conclude with a range of observational tests of overluminous SNe Ia which will either support or strongly constrain the single-degenerate scenario. 

\end {abstract}


\section {Introduction}

Type Ia supernovae (SNe Ia) play an important role as standardizable candles for cosmology \citep {Schmidt_1998, Riess_1998, Perlmutter_1999}. Yet, the nature of their stellar progenitors remains elusive \citep {maozetal14}. Several leading models have been advanced. In the classic single-degenerate channel, a hydrogen-rich main sequence or red giant star accretes matter onto a white dwarf companion, until the companion nears or exceeds the Chandrasekhar mass and explodes \citep {Whelan_Iben_1973}. In contrast, the double-degenerate channel posits that double C/O white dwarf systems, brought together by gravitational waves, merge and lead to SNe Ia. Additionally, WDs accreting helium from either a non-degenerate or WD companion may also give rise to a SN Ia through the double-detonation channel \citep {nomoto82}. Recent observational evidence
%, ranging from the delay-time distribution (DTD) \citep {totanietal08, grauretal11, maozbadenes10, maozetal10, maozetal12, grauretal14}, the absence of H in nebular phase \citep {leonard07},  and the absence of ex-companions 
has begun to favor  double-degenerates, or possibly double-detonations, as the origin for the majority of SNe Ia.  On the other hand, additional evidence -- ranging from bimodality in spectral features \citep {benettietal05},  circumstellar material \citep  {dildayetal12}, and the ejected mass-$^{56}$Ni relation  \citep {scalzoetal14}, as well as  stable iron peak-element nucleosynthesis \citep {seitenzahletal13, yamaguchietal15} -- has hinted that multiple channels may contribute to the overall SNe Ia rate.

Therefore, even if double-degenerate or double-detonation supernovae do account for the majority of normal SNe, a number of major questions remain unresolved. Specifically, what is the fate of rapidly-accreting white dwarfs in the single-degenerate channel once they reach the Chandrasekhar mass? If the single-degenerate channel does give rise to SNe Ia, how can we reconcile the existence of single-degenerate supernovae with the general absence of observational evidence in their support? Perhaps most crucially, if there are indeed multiple channels producing SNe Ia, what are the specific observational characteristics which would enable us to separate the population of single-degenerate SNe Ia from the population of SNe Ia as a whole, and most directly confront observations with theory?

To address these questions, it is important to recall that SNa Ia are standardizable candles, but  exhibit a range of diverse outcomes, from failed SN 2002cx-like events, through normal SNe Ia, to overluminous SN 1991T-like SNe Ia \citep {filippenkoetal92}, and superluminous supernovae such as 2003fg \citep {howelletal06} or 2009dc  \citep {yamnakaetal09, silvermanetal11}.  In this paper, we will explore the possible connection of single-degenerate SNe Ia to the population of overluminous SNe Ia. Overluminous SNe Ia  account for 4 - 20\% of all SNe Ia \citep {foleyetal09, lietal01}, and yield a significantly larger amount of $^{56}$Ni than normal SNe Ia (roughly 1 $M_{\odot}$, as opposed to $0.5 - 0.8 M_{\odot}$). 

The rates and high $^{56}$Ni yields of overluminous SNe Ia  are in tension with both double-detonation and double-degenerate merger models of SNe Ia, since both fundamentally rely upon sub-Chandrasekhar mass WDs. The prompt detonation of cold, sub-Chandrasekhar C/O WDs up to $1.15 M_{\odot}$ produce only up to $.81 M_{\odot}$ of $^{56}$Ni, which is at the upper end of the range of $^{56}$Ni produced in normal SNe Ia \citep {simetal10}. Therefore, the production of the amounts of  $^{56}$Ni required for overluminous events  in either the  violent double-degenerate or double-detonation scenario requires massive C/O primaries in excess of $1.2 M_{\odot}$, and approaching the Chandrasekhar mass, which both observation and modeling suggest are very rare in both channels \citep {badenesmaoz12, ruiteretal11, ruiteretal13}. It may be possible for two sub-Chandrasekhar white dwarfs to merge into a near-Chandrasekhar mass white dwarf on a viscous evolutionary timescale and produce an overluminous SN Ia \citep {vankerkwijketal10, piroetal14}, though it is remains unclear precisely whether such mergers will detonate \citep {shenetal12, schwabetal12, jietal13}.
%Furthermore, while calculations of the violent merger scenario in very massive binaries $1.2$ + $1.1$ $M_{\odot}$ C/O WDs do yield $.95 M_{\odot}$ of $^{56}$Ni, 
%%RTF: (CITE PAKMOR THESIS)
%it remains unclear whether their spectral and light curve properties are consistent with observations of SN 1991T-like SNe Ia. 
Furthermore, in the context of double-detonations, models suggest that even the minimal mass of a thin detonating shell of helium will in turn detonate the underlying C/O WD core \citep {finketal10}. Consequently, He-accreting massive C/O WDs may tend to detonate easily, with little opportunity to accrete the requisite mass to explain overluminous SNe Ia.
 
 %In the context of violent mergers, which are the most throughly-explored mechanism for double-degenerate SNe Ia to date, the detonation proceeds on a dynamical timescale. As a result, the density profile of the primary remains unchanged during the rapid merger and detonation phase of violent mergers. Since the densities of the tidally-disrupted secondary are typically too low to reach NSE conditions, the amount of  $^{56}$Ni  generated during the violent merger mechanism is essentially established by the mass of the primary. In these conditions, even for relatively massive mergers, of $1.1$ + $1.0$ $M_{\odot}$ C/O WDs, the resulting  $^{56}$Ni nucleosynthetic yield is typically $\sim 0.6 M_{\odot}$ \citep {pakmoretal12}, also characteristic of normal brightness SN Ia. Only in extremely massive binaries $1.2$ + $1.1$ $M_{\odot}$ C/O WDs does the violent merger scenario produce $.95 M_{\odot}$ of $^{56}$Ni (CITE PAKMOR PHD THESIS). Such very massive C/O WD primaries in excess of $1.2 M_{\odot}$ do indeed appear to produce luminosities consistent with SN 1991T-like overluminous events, though they occur in roughly $5 \%$ of all C/O WD merging binaries \citep {ruiteretal13}. Because the violent merger scenario additionally requires a high-mass ratio of C/O WDs, overluminous SNe are expected in less than $1\%$ of the double-degenerate channel. 
 
In contrast, overluminous supernovae have long been understood to be the natural consequence of the detonation of Chandrasekhar-mass WDs \citep {arnett69}. However, until recently,  conventional lore has held that the near-uniformity of normal SNe Ia strongly favored the single-degenerate channel as the progenitors of normal SNe Ia. For decades, theorists have worked to reconcile the class of normal SNe Ia with the relatively large amount of $^{56}$Ni produced by the explosion of cold Chandrasekhar-mass white dwarfs. The most promising theoretical framework for over two decades has been the deflagration-to-detonation transition (DDT), in which a turbulent flame ignited within the convective core of a Chandrasekhar-mass white dwarf leads to slow, subsonic burning, and a pre-expansion of the white dwarf itself \citep {khokhlov91, gamezoetal05}. The DDT conjectures that the flame undergoes a subsequent transition to a detonation, with the material in the pre-expanding star producing a normal Ia event. However, as we explain here, recent theoretical work on the convective simmering phase within Chandrasekhar-mass white dwarfs is inconsistent with the large amount of burning required by the standard DDT model for normal SNe Ia during the deflagration phase in the majority of ignitions. 

In this paper, we present the case that single-degenerate SNe Ia generally lack a vigorous deflagration phase,  and preferentially result in overluminous SNe Ia.   This hypothesis  is motivated by the need to reconcile numerous recent observational results, ranging from the DTD \citep {totanietal08, grauretal11, maozbadenes10, maozetal10, maozetal12, grauretal14}, the absence of evidence for companions  \citep {lietal11, bloometal12} and ex-companions  \citep {maozmannuccietal08, gonzalezhernandez12, kerzendorfetal12, schaeferpagnotta12, edwardsetal12, kerzendorf_etal_2014}, the absence of H in the nebular phase \citep {leonard07}, and the absence of radio and X-ray emission \citep {lietal11, brownetal12, horeshetal12, bloometal12, marguttietal12, chomiuketal12} around peak brightness,  with recent theoretical advances. In particular, recent theoretical work has, for the first time, clarified the inherently stochastic nature of carbon ignition in the convective cores of Chandrasekhar-mass white dwarfs. The literature of explosion models has focused primarily either upon precisely-centered and significantly off-centered ignitions, with widely ranging outcomes dependent upon the ignition -- from subluminous 2002cx-like events \citep {jordanetal12a, kromeretal13}, though overluminous SNe Ia events \citep {plewaetal04, jordanetal08, meakinetal09}. Here, we develop an analytic criterion that separates between buoyancy-driven and central ignitions. Criteria for central ignition have been explored previously in the literature \citep {zingaledursi07, aspdenetal11, maloneetal14} using numerical solutions to semi-analytic models and fully multidimensional simulations. Here, we provide a new, elementary derivation of the critical length scale within which central ignitions may arise, which highlights the essential competition  between the physical processes of burning and buoyancy.  We further demonstrate that the majority of ignition events within the single-degenerate channel lead to buoyancy-driven turbulent deflagration, with little deflagration energy release and pre-expansion, and argue these will be followed by detonation and consequently an overluminous SNe Ia. We place bounds on the number of normal and subluminous SNe Ia events originating from the single-degenerate channel, and demonstrate that these can only account for a small fraction of all single-degenerate events. We argue that normal and subluminous events must primarily arise through the double-degenerate and helium-donor channels.

These findings clarify a number of long-standing problems, including the rate problem of the  single-degenerate channel, the absence of observed stellar progenitors and ex-companions, and the absence of hydrogen in the nebular phase of normal supernovae. They also connect the particular class of overrluminous SNe Ia to stellar progenitors, including recurrent novae and supersoft X-ray sources, and strengthen the association of single-degenerate channel SNe Ia to the subclass of SN Ia-CSM events like PTF11kx \citep {dildayetal12}.  Additionally, they shed additional light on why overluminous supernovae primarily originate in late-type galaxies. Our results offer guidance to observers to look towards overluminous events for the most likely occurrence of evidence either to support or constrain the single-degenerate channel.
% including early-time light curve distortions, hydrogen H$\alpha$ emission and ex-companions.

%In this paper, we weigh the preponderance of evidence from observation and theory, and suggest that the weight of evidence, both observational and theo. This is certainly not the first attempt to capture a unified view of SNe Ia. Recent work by other researchers has focused upon the double-degenerate channel (Pakmor et al, 2013), and the single-degenerate channel (Wheeler, 2012).  However, recent progress on both normal and subluminous supernova challenge models focusing primarily on individual channels as a unifying framework for all SNe Ia events. For instance, multi-wavelength observations derived from the normal SN 2011fe put strong constraints on the single-degenerate channel for that event. Other observations of the subluminous, low-velocity SN 2012Z point towards a helium main sequence donor, and (if confirmed) will rule out double-degenerate mergers as the sole channel for both normal and subluminous events. Consequently, a re-evaluation of all available evidence is timely.

%In section \S \ref {sec:observational}, we present a summary of the key observational constraints derived for normal and overluminous Ia events.
In \S \ref {sec:theory}, we review the theory of single-degenerate Chandrasekhar-mass white dwarfs, from ignition through detonation. In \S \ref {sec:ignition}, we delve into the ignition problem in depth, and develop an analytic criterion that  separates between buoyancy-driven and central ignitions. In \S \ref {sec:rates}, we compare the rates both of supersoft X-ray sources (SSSs), as well as  binary population synthesis models, against those of overluminous Ia events. In \S \ref  {sec:implications}, we conclude with the observational implications of this  work, and suggest specific observations which can either directly support or constrain the single-degenerate channel as a primary contributor to overluminous SNe Ia events.

\section {Single-Degenerate Channel}
\label {sec:theory}

The single-degenerate channel has been the most thoroughly-explored SNe Ia model framework over the last several decades \citep {Whelan_Iben_1973}. It begins with an intermediate mass primary star between 3 $M_{\odot}$ - 8 $M_{\odot}$ and a companion. The primary evolves more rapidly, and transfers mass to the companion, leading to a common envelope phase of evolution. The primary forms a 0.6 - 1.2 $M_{\odot}$ C/O WD, which can then reach the Chandrasekhar mass by hydrogen-rich mass transfer from the non-degenerate companion. Modeling this sequence of events using theoretical binary population synthesis (BPS) codes demonstrates the broad feasibility of the channel, although the predicted rates  are subject to considerable uncertainties, which we will review in \S \ref {sec:rates}. Observationally, the accreting white dwarfs are visible as recurrent novae and supersoft X-ray sources. The more rapidly-accreting supersoft sources accumulate mass up to the Chandrasekhar mass limit, and contrary to theoretical expectations, recent observational evidence of the mass functions of pre-CV and post-CV WDs seems to suggest that recurrent novae may also accumulate mass \citep  {zorotovicetal11}. 

The central conditions of accreting white dwarfs  reach carbon ignition as the white dwarf approaches the Chandrasekhar mass limit \citep {nomotoetal84}. A variety of explosion models have been considered, ranging from pure deflagrations \citep { ropkeetal07b, Maetal13}, through deflagration-to-detonation transitions (DDTs) \citep {gamezoetal05, ropkeniemeyer07}, and gravitationally-confined detonations (GCDs) \citep {plewaetal04, townsleyetal07, jordanetal08, meakinetal09, jordanetal12b, jordanetal12a}.   

It has slowly become clear that, depending on the nature of ignition, the single-degenerate channel predicts a wide range of deflagration energy yields.  The greater the deflagration energy release, the greater the pre-expansion, and the lesser the yield of $^{56}$Ni during the following  detonation phase. Thus, the single-degenerate channel predicts a range of outcomes, from  overluminous events like SN 1991T \citep {jordanetal08}, through  normal Ia events \citep {gamezoetal05, jordanetal12b},  and even,  with highly vigorous deflagration burning and in the absence of any detonation, subluminous SN 2002cx-like events \citep {jordanetal12a, kromeretal13}. The energetic yield during the deflagration phase is directly tied to the ignition problem, which until very recently was considered a largely unsolved problem, with theorists addressing the uncertainty by widely varying the number and offset location of ignition points in multidimensional models of the deflagration and detonation phases. However, a monumental computational campaign, primarily by Zingale, Woosley, Bell and collaborators \citep {zingaleetal11, nonakaetal11, maloneetal14} has revealed that the simmering phase typically yields a single ignition point, with a distribution of offset ignition locations, and an expectation value of $\sim 50$ km. 

The distribution of offset locations reflects the inherently stochastic nature of turbulent convection. This irreducible stochasticity of thermonuclear ignition directly results in an inherently probabilistic range of outcomes for the single degenerate channel. Sufficiently large offset radii lead to buoyancy-driven ignition and deflagration, but precisely how offset must an ignition point be to lead to this outcome? This is the key question which we ask and address in the next section, where we quantify the probability of a central ignition by deriving a characteristic length scale separating buoyancy-driven ignitions from central ignition events.

\section {Ignition and Detonation}
\label{sec:ignition}

\subsection {Central and Offset Ignition}

A single buoyancy-driven ignition point burns a small fraction of the white dwarf during the deflagration phase \citep {plewaetal04, townsleyetal07, ropkeetal07a, jordanetal08,meakinetal09, maloneetal14}, and consequently leads to small amounts of pre-expansion, consistent with an overluminous SNe Ia event. In contrast, central ignition points linger near the central region of the star,  leading to greater pre-expansion,  typically producing  $\sim 0.5 M_{\odot}$ of $^{56}$Ni burnt during the deflagration phase \citep {nomotoetal84, gamezoetal05, Maetal13}. This greater amount of burning during the deflagration phase may yield a subluminous SNe Ia by itself \citep {jordanetal12b, kromeretal13}, or a normal SNe if followed by a detonation, in either the context of a DDT or GCD \citep {gamezoetal05, meakinetal09, jordanetal12a}.  Hence, for single-bubble ignitions, a key question is which ignition radii lead to buoyany-driven ignitions, and which lead to central ignitions. The answer to this question is crucially connected to the amount of pre-expansion which the WD experiences during the deflagration phase, which in turn determines whether a DDT or GCD is possible, as we will see in section \ref {sec:detonation}.

We now develop an analytic criterion which demarcates the boundary between buoyancy-driven and central ignitions.  We assume a single ignition bubble, with offset radius $r_0$ from the center of the white dwarf, and radius $R$ -- see figure 1. We assume that the ignition size is small in comparison to the offset radius -- $R \ll r_0$.  A crucial critical length scale in the development of the flame bubble is the fire-polishing scale $\lambda_{\rm fp} = 4 \pi S_l^2 / A g$. Perturbations to the flame bubble on length scales below $\lambda_{\rm fp}$ are polished away by the action of the flame, whereas larger-scale perturbations are subject to the Rayleigh-Taylor instability \citep {timmeswoosley92}. For typical ignition conditions, the bubble radius $R$ is much less than the fire-polishing scale $\lambda_{\rm fp}$ and as a result, is necessarily spherical upon ignition.  Furthermore, the bubble radius is also much smaller than the integral length scale $L \sim 100$ km of the convective turbulent background. Consequently, the turbulence associated with convection is also negligible on the scale of the bubble during its initial growth. 

While the neglect of both the non-spherical distortion of the bubble and of turbulence is an excellent approximation initially, they begin to break down as the bubble burns outward \citep {maloneetal14}. However, for the purpose of determining which ignition points are buoyancy-driven, we will be interested in only the initial growth of the bubble, during which both approximations continue to hold valid throughout. Furthermore, numerical simulations demonstrate that the bubbles are typically ignited within a convective plume with a net radial velocity, though the magnitude of the typical plume velocity ($\sim 20$ km/s) is relatively small in comparison to the laminar flame speed $S_l \sim 100$ km/s, and may also be neglected to a first approximation. Further, we demonstrate below that the neglect of the plume velocity places a conservative  upper-bound on the possible central ignition events. Therefore, we will consider the initial growth of the bubble on a purely stationary background.

\begin{figure}
\begin{center}
\includegraphics[width=0.7\columnwidth]{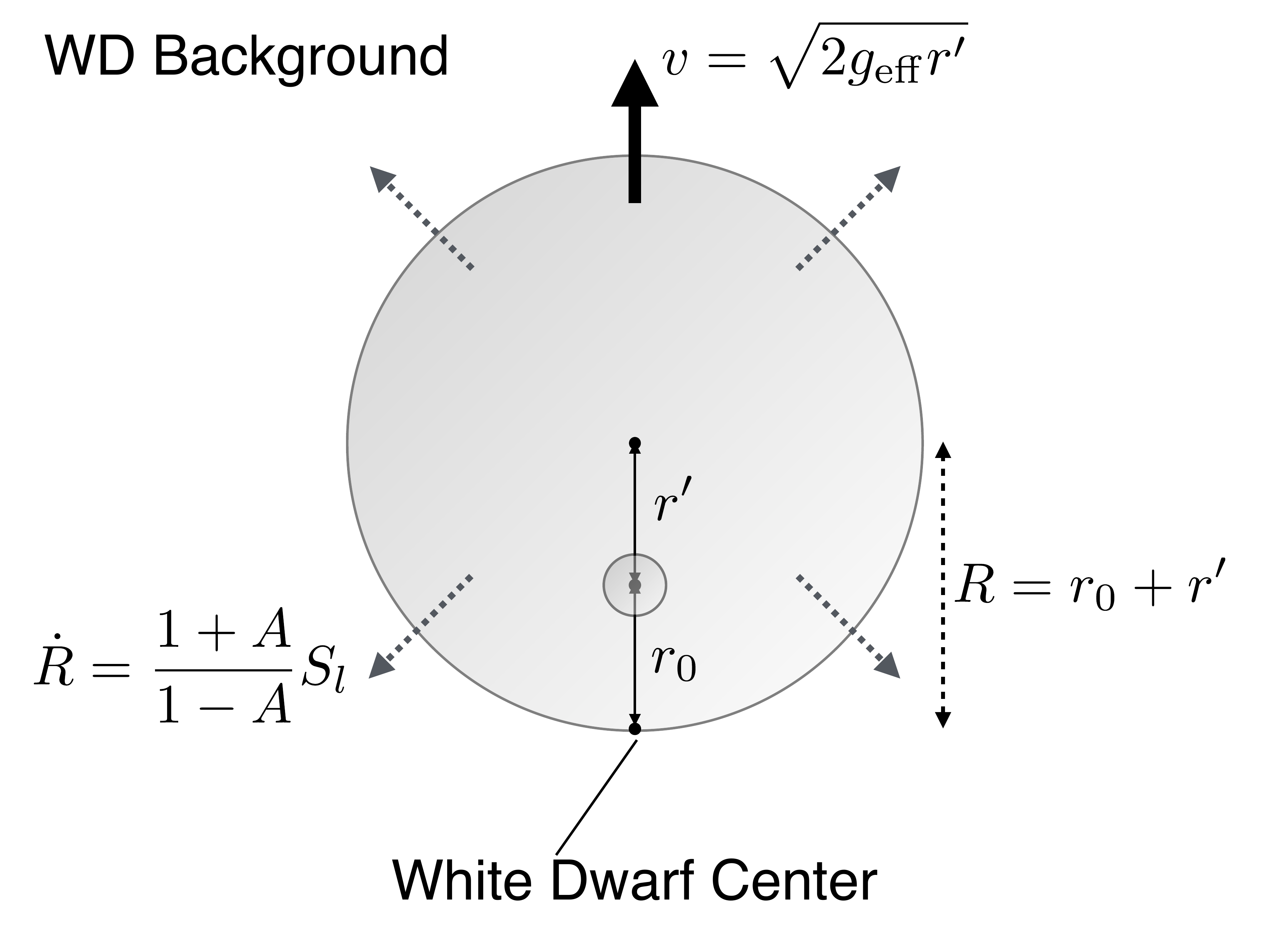}
\caption{A schematic diagram depicting the geometry of the critical case of a bubble, ignited at a distance $r_0$ from the center of the star, which just barely begins to burn through the center of the star in the time required for it to buoyantly rise an additional vertical distance $r'$. As shown, the radius $R$ of the critical case bubble at the instant it begins to burn through the center of the star is $R = r_0 + r'$.  }
\label {fig:bubble_diagram}
\end{center}
\end{figure}

%Ignition points which burn through the central region of the white dwarf before experiencing significant  buoyancy, lead to a large deflagration energy release. In contrast, ignition points which do not meet this criterion are dominated by buoyancy, and typically burn only a small amount during the deflagration phase.  

The criterion for central ignition is simply expressed; the bubble radius $R (t)$ must exceed the offset radius $r (t)$ at some time $t$ -- that is, $R (t) > r (t)$. A central ignition burns through the center of the star prior to buoyantly rising an additional vertical distance $r'$.  In the appendix material, we consider the combination of buoyancy,  drag forces, and the added mass effect on a self-similar spherical flame bubble that develops from a point-like ignition at $t = 0$  -- that is, $R (t = 0) = 0$. There, we demonstrate that the combination of these effects result in an effective gravitational acceleration $g_{\rm eff}$. That is, the leading terms of a power-series solution of the  equation of motion of the bubble, an ordinary differential equation describing the radial location $r$, equation (A8), are given by $r(t) = r_0 + (g_{\rm eff} / 2) t^2$ at early times after ignition, with $g_{\rm eff}$ is given by
\begin {equation}
g_{\rm eff}  = {A g \over ( 3 - A)}  
\end {equation}
Here $A$ is the Atwood number, a dimensionless quantity describing the density contrast across the bubble interface :
\begin {equation}
A = {\rho_f - \rho_a \over \rho_a + \rho_f}
\end {equation}
where $\rho_f$ is the density of the fuel, and $\rho_a$ is the density of ash, and the effective gravitational acceleration $g_{\rm eff}$ is taken at the initial radius $r_0$. The density of fuel and ash are similar under carbon burning at central densities typical of Chandrasekhar-mass white dwarfs, leading to small Atwood number values $A \simeq 0.09$ \citep {timmeswoosley92}. As shown in the appendix, in equation (A5), in the absence of both drag and the added mass effect, the effective buoyant acceleration would simply be

\begin {equation}
g_{\rm eff} = g  \frac{\Delta \rho}{\rho}
\end {equation}
where ${\Delta \rho}/ {\rho} = 2A / (1+A)$. Consequently, the effective gravity for a low Atwood number bubble ($A \ll 1$) at early times, including the effects of drag and added mass, is very nearly $1/6$ of the buoyancy-driven value that would be obtained in the absence of both drag and the added-mass effect. 

%In the time interval $t$ that a bubble rises through a vertical distance $r_0$ equal to its ignition radius, it burns radially outward by a distance $(1 + A) / (1 - A) S_l t$, where $S_l$ is the laminar flame speed.  The time $t$ required for the bubble to burn through the center of the star is then simply
%
%\begin {equation}
%t = \left( {1 - A \over 1 + A}   \right) {2 r_0 \over S_l}  
%\end {equation}
%
%Consequently, a necessary and sufficient condition for a buoyancy-driven ignition is that it rises through the vertical distance $r_0$ prior to burning through the center of the star. 

The bubble achieves a buoyant velocity $v = \sqrt {2 g_{\rm eff} r'}$ after rising the radial distance $r'$.  In the critical case, the buoyant velocity $v$ will exactly equal the radial expansion velocity of the bubble $[(1 + A) / (1 - A)] S_l$ (equation A1), and the bubble just barely burns through the center as it reaches the offset radius $r_0 + r'$.  Hence,

\begin {equation}
r' = \left( { 1 + A \over 1 - A} \right)^2 {S_l^2 \over 2 g_{\rm eff} }
\end {equation}

Furthermore, the critical condition that the radius $R$ of the bubble reaches the center of the star in this same time, $t = \sqrt {2 r' / g_{\rm eff} }$ yields

\begin {equation}
R = r_0 + r' = \left( { 1 + A \over 1 - A} \right)^2 {S_l^2 \over g_{\rm eff} }
\end {equation}
Consequently, in the critical case, the additional distance the bubble must rise to burn through the center of the star is precisely equal to its offset radius: $r' = r_0$. Therefore, a buoyancy-driven ignition must exceed a critical offset radius $r_{\rm crit}$

\begin {equation}
r_0 >   r_{\rm crit} = \left( {1 + A \over 1 - A}   \right)^2 {S_l^2  \over2  g_{\rm eff}}  =   {1 \over 8 \pi}  \left( {1 + A \over 1 - A}   \right)^2 \lambda_{\rm fp} 
\label {eqn:criticalradius1}
\end {equation}
This result demonstrates that the critical offset radius for a buoyancy-driven ignition event is proportional to, but significantly smaller than the fire-polishing scale, with the laminar flame speed and effective gravity taken at the ignition location. 
%%The buoyant velocity $v = \sqrt {2 g_{\rm eff} r_{\rm crit}} = (1 + A) (1 - A)^{-1} S_l \simeq S_l$ for low Atwood number bubbles at the critical offset radius. Consequently, because the buoyant velocity continues to accelerate as the bubble moves outward even as the laminar flame speed drops, bubbles which fail to burn through the center by the critical offset radius $r_{\rm crit}$ continue to rise away from the star without ever crossing the center.

The competition between laminar burning and buoyancy is intrinsically linked through the fire-polishing scale $\lambda_{\rm fp}$. In particular, because the fire-polishing scale is inversely proportional to the effective gravitational acceleration, near the center of the star, where $g_{\rm eff} \rightarrow 0$, the laminar burning timescale is shorter than the buoyancy timescale of the bubble, and bubbles ignited there must burn through the center of the star. Conversely, at larger offset radii, as the effective gravitational acceleration increases,  the fire-polishing scale $\lambda_{\rm fp}$ drops, and the buoyancy timescale is shorter than the laminar burning timescale. The critical radius $r_{\rm crit}$ defined by equation \ref {eqn:criticalradius1} establishes the boundary between these two regimes.

The local effective gravitational acceleration may be expressed as $g_{\rm eff} = (4 \pi / 3) G_{\rm eff}  \rho_c r_0$, with $G_{\rm eff} = [A / (3 - A)] G$ the effective gravitational coupling for the bubble:  
\begin {equation}
r_{\rm crit} = \left( {1 + A \over 1 -  A}  \right) \sqrt  {3 \over 8 \pi  G_{\rm eff}   \rho_c}  S_l 
\end {equation}
Further insight into the physics of the buoyancy-driven regime can be gathered by noting that buoyancy-driven ignition occurs when the initial offset radius $r_0$ exceeds a value which is proportional to the product of the laminar flame speed and the central buoyant dynamical time :
%
%\begin {equation}
%_0 >  \left( {1 + A \over 1 - A}  \right) {2  \over \pi} S_l t_{\rm dyn} = 25\ {\rm km} \left (S_l \over 90\ {\rm km/s}  \right) \left (.09 \over A  \right)^{1/2}  \left (2.2 \times 10^9\ {\rm gm/cm^3} \over \rho_c  \right)^{1/2} 
%\end {equation}
%
%\begin {multiline}
%r_0 >  \left( {1 + A \over 1 - A}  \right) {2  \over \pi} S_l t_{\rm dyn} = \\25\ {\rm km} \left (S_l \over 90\ {\rm km/s}  \right) \left (.09 \over A  \right)^{1/2}  \left (2.2 \times 10^9\ {\rm gm/cm^3} \over \rho_c  \right)^{1/2} 
%\end {multiline}

\begin{align*}
r_{\rm crit}  &=  {2  \over \pi} \left( {1 + A \over 1 - A}  \right)  S_l t_{\rm dyn}\\  &= 
19\ {\rm km} \left (S_l \over 100\ {\rm km/s}  \right)  \left (.09 \over A  \right)^{1/2}  \left (2.2 \times 10^9\ {\rm gm/cm^3} \over \rho_c  \right)^{1/2} 
\end{align*}
Here the dynamical time is defined as the buoyancy dynamical timescale $t_{\rm dyn} = \sqrt {3 \pi / (32 G_{\rm eff}  \rho)}$, based upon the effective gravitational acceleration, and the the fiducial scales are determined for a 50/50 C/O WD \citep {timmeswoosley92}.
%Moreover,  the time $2 r_{\rm crit} / S_l$ required for a bubble ignited at the critical transition radius to both burn through the center of the star and reach a buoyant offset radius $2 r_{\rm crit}$ is proportional to the central buoyant dynamical time, and is independent of the laminar flame speed.

%%
%\begin {align*}
%{2 r_{\rm crit}  \over S_l} &=  {4  \over \pi} \left( {1 + A \over 1 - A}  \right)  t_{\rm dyn}\\
%                                       &=  .55\ {\rm s}  \left (2.2 \times 10^9\ {\rm gm/cm^3} \over \rho_c  \right)^{1/2} 
%\end {align*}
%

The critical radius $r_{\rm crit}$ is very close to the center of the WD, and highlights the importance of an analytic derivation. In figure 2, we plot the analytic solution trajectory of bubble radius versus offset radius for a buoyancy-driven ignition, a central ignition, and the critical ignition case. For comparison, we also overplot a full numerical integration of equations (A1) and (A3) on a fully-stratified white dwarf background, for which we determined a critical ignition radius $r_{\rm crit} = 24.1$ km, and show very good agreement with the critical solution trajectory. Large-scale three-dimensional full-star numerical simulations typically have spatial resolutions ranging from 4 - 8 km, and therefore only marginally resolve the critical transition radius at best. Some comparisons, can however, be made between our analytic result and those obtained from previous analytic modeling and three-dimensional numerical simulations. \citet {aspdenetal11} developed a semi-analytic model very similar to the analytic one presented in the appendix, though including both an entrainment term for the bubble growth as well as two free parameters which were then fit to data from three-dimensional models of the ignition and early evolution of the flame bubble, and then solved numerically on a full-star background. \citet {aspdenetal11} present evidence for a transition to central ignition somewhat interior to $r_0 = 16$ km, though it should be noted that they used a bubble with an initial radius of 10.2 km, which is comparable to the critical radius, and adopted a somewhat higher central density  $2.55 \times 10^9$ g cm$^{-3}$,  than assumed here. \citet {zingaledursi07} numerically find $r_{\rm crit} \sim 23.5$ km, using small initial bubble radii like our self-similar solution, and in very good agreement with our own numerical solution. However, they assume a higher central density of $2.6 \times 10^9$ g cm$^{-3}$  and employ a fit for the laminar flame speed that is modified to improve its accuracy at low densities compared to the fit from \citet {timmeswoosley92} that we use, in addition to assuming a lower carbon mass fraction of 0.3.  A more recent study by \citet {maloneetal14} finds a critical radius of $\sim 10$ km over a range of point-like initial bubbles. Similarly, a variety of other two and three-dimensional full-star simulations find a significant and sharp enhancement in the burning during the deflagration phase for $r_0 \lesssim 20$ km --- see e.g. \citet {plewa07} and \cite{meakinetal09}, though the resolutions in these models is relatively coarse, and begin with bubbles with radial extents comparable to the critical radius. Overall, our analytic result is, however, consistent with these previous findings.

\begin{figure}
\begin{center}
\includegraphics[width=1.0\columnwidth]{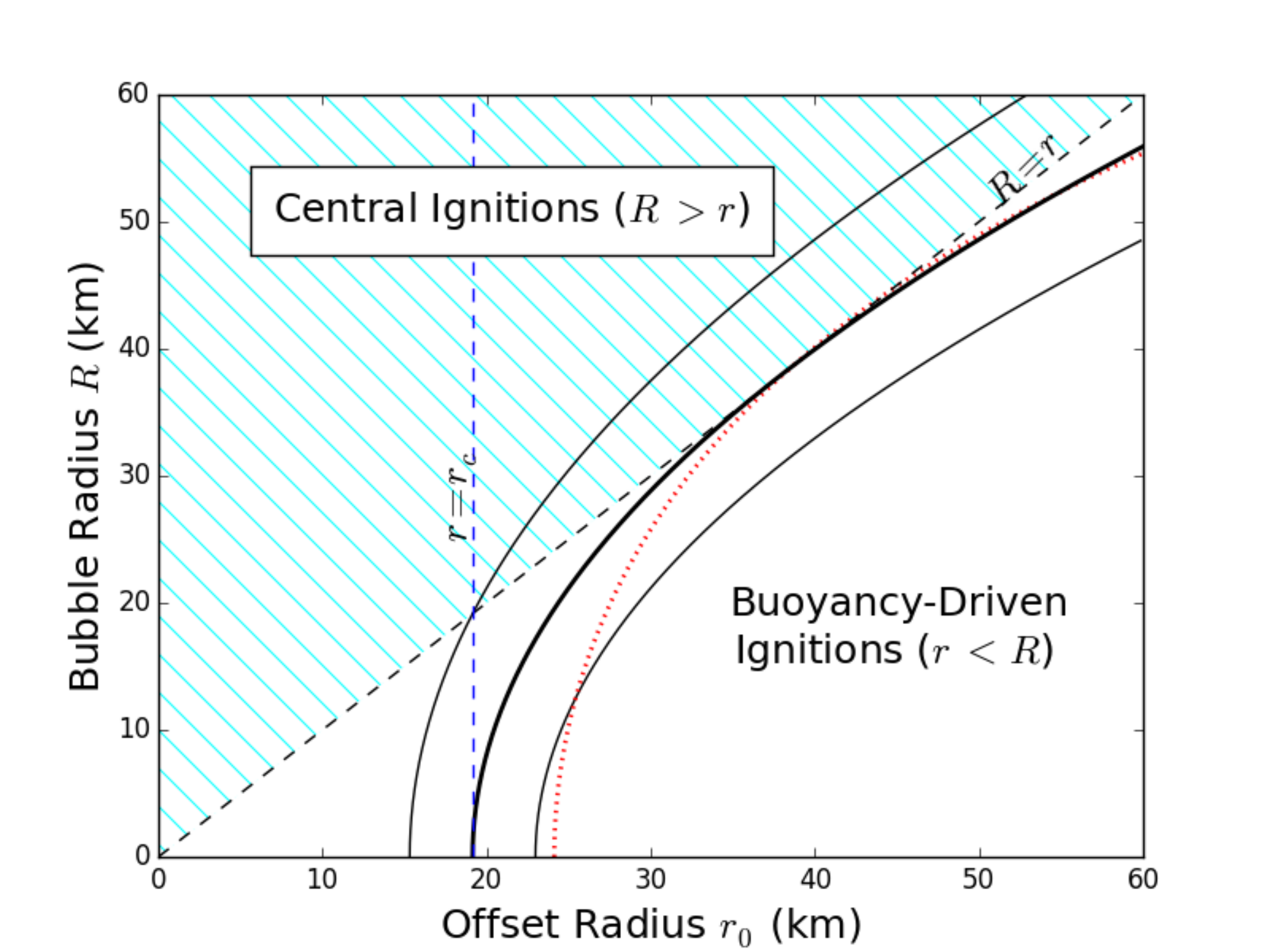}
\caption{A plot of bubble radius $R$ versus offset radius $r_0$. The dashed line $R = r$ is the critical case separating central ignitions from buoyancy-driven ignitions. The hatched region of the figure marks central ignitions, with $R > r$; buoyancy-driven ignitions lie in the portion of the figure with $r < R$. The vertical dashed line shows the critical offset radius, and the solid lines denote the trajectories of the analytic solution described in the text, with the thick solid line showing the critical case of a bubble which barely burns through the center of the star. For comparison, a full numerical integration of a critical bubble in a fully-stratified WD, with an initial critical offset $r_{\rm cit} = 24.1$ km is shown in the dotted curve.}
\label {fig:figure_2}
\end{center}
\end{figure}

Numerical simulations of the convective phase of Chandrasekhar-mass white dwarfs have revealed the distribution of hot spots within the core. \citet {zingaleetal11} and \citet {nonakaetal11} have examined the distribution of hot spots leading up to ignition. Due the strongly nonlinear character of the ignition problem, only one of these hot spots ignites and runs away. We note that, due to the significant computational cost of these three-dimensional numerical simulations of convection and ignition, only a single progenitor model has been examined in detail to date, and additional models will be needed to further elucidate the nature of the ignitions. However, we consider this distribution of hot spots as representative of the temperature distribution of all possible ignition points within the white dwarf, and therefore representative of the statistical distribution which would be obtained for the final ignition points over an ensemble of simulations. 

Conservatively, from direct inspection of their figure 7, we find that no greater than $2 \pm 1\%$ of the ignition hot spots considered by \citet {nonakaetal11}  burn through the center of the star, and lead to central ignitions, with a commensurately-larger energy release during the deflagration phase.  The remaining hot spots will lead to buoyancy-driven ignition events. We emphasize that these estimates are conservative upper-bounds on the likelihood of a central ignition, due to our neglect of the generally upward plume velocities which the hot spots are ignited within, and to the enhanced effect of buoyancy in the numerical simulations over analytic models \citep {maloneetal14}.  Thus, the overwhelmingly likely outcome of ignition in a Chandrasekhar-mass white dwarf is a buoyancy-driven event.  As we demonstrate in the next section, these buoyancy-driven ignition events are  in turn generally expected to lead to overluminous SNe Ia.

\subsection {Detonation}
\label {sec:detonation}

Pure deflagration models  \citep { ropkeetal07b, Maetal13} have numerous deficiencies when compared against observations, including significant amounts of unburned carbon and oxygen, which can be rectified if the white dwarf undergoes a subsequent detonation. Hence, for many years, the leading single-degenerate channel model has been the deflagration-to-detonation transition (DDT) \citep {khokhlov91, gamezoetal05, ropkeniemeyer07}. However, the DDT model crucially rests on a significant amount of nuclear burning during the deflagration phase, prior to the subsequent detonation, in order to pre-expand the WD and achieve a nucleosynthetic yield of $^{56}$Ni typical of normal brightness SNe Ia. {\it A key consequence of the central ignition criterion, equation (\ref {eqn:criticalradius1}), is that such a vigorous deflagration phase is highly unlikely in the majority of single-degenerate SNe Ia ignition events.} Instead, the overwhelmingly-likely outcome of ignition is a buoyancy-driven flame bubble with a relatively small deflagration energy yield.

The subsequent evolution of a buoyancy-driven flame bubble has been explored in numerous calculations in the literature. In a gravitationally-confined detonation (GCD), the small amount of nuclear energy released during the deflagration phase leads to a bound white dwarf, and the hot bubble breaks out of the bound white dwarf surface.  In a GCD, the white dwarf undergoes a detonation as the hot ash remaining from the bubble is ram-driven into high densities \citep {plewaetal04, townsleyetal07, jordanetal08}. We note that these findings were not initially universally agreed-upon, with another group using an enhanced subgrid model for turbulent combustion found a much greater degree of pre-expansion, leading to a detonation in 2D but not 3D \citep {ropkeetal07a}. However, more recent work, both semi-analytic and numerical, has demonstrated the deflagration energy release  is established by large-scale hydrodynamic entrainment of the bubble, and is less sensitive to the turbulent flame model adopted \citep {aspdenetal11}. Additionally, a third group \citep {maloneetal14} modeling the convective phase though ignition and bubble breakout has obtained deflagration results very similar to that of the classic GCD model of \citet {jordanetal08}, using a completely independent code.

There are several caveats to consider when evaluating this theoretical work. Firstly, our physical and computational description of turbulent deflagration is still limited. Much of the burning during the deflagration phase occurs on length scales unresolved in multi-dimensional numerical simulations. Several approaches have been advanced to address with this issue. While there are some differences in the amount of burning in the deflagration predicted by these approaches, they generally agree in that the amount of nuclear energy released for a single bubble ignition is small in comparison to the gravitational binding energy of the WD. Only a significant revision of our basic understanding of turbulent combustion would fundamentally alter this outcome. Moreover, recent analytic and semi-analytic work on burning bubbles has clarified that the large-scale hydrodynamical entrainment plays a more important role than the choice of the turbulent flame speed \citep {aspdenetal11}. 

Secondly, three-dimensional simulations of both the convective phase and the subsequent deflagration and detonation phases typically rely upon a standard C/O WD model, with 50/50 C/O composition, and masses 1.36 - 1.38 $M_{\odot}$. However, the deflagration phase is sensitive to the value chosen for this mass, primarily through the nonlinear dependence of the laminar flame speed to the density, leading to commensurately greater flame polishing at the higher central densities for WD progenitors approaching the Chandrasekhar mass \citep {kruegeretal12, lambetal14}. Higher laminar flame speeds polish the flame bubble surface, and consequently produce less burning than a lower laminar flame speed. Even a very small relative change in the WD mass, of 1\% or less, at ignition can produce a significant change in the energy yield in the deflagration phase. Thus it is possible to produce a range of outcomes, even including normal SNe Ia, by fine-tuning of the chosen progenitor WD mass very close to the Chandrasekhar mass. We view such a fine-tuning as unlikely, since it would tend to produce a relatively large rate of normal single-degenerate SNe Ia, in tension with observations. However, this is an important issue which needs to be understood better, since it may represent the dominant theoretical uncertainty of theoretical modeling of the single-degenerate channel. 

Additionally, a buoyancy-driven flame bubble may undergo a transition to detonation before it even breaks out from the white dwarf surface. Such a transition to detonation would differ significantly from the classic DDT scenario, however, due to the lower deflagration energy release, with the subsequent outcome still inevitably an overluminous event. Lastly, the first bubble may not produce an effective explosion, and may instead be followed by a series of successive bubble ignitions, only one of which is required to detonate the white dwarf, as speculated recently by \citet {maloneetal14}.

In summary, the bulk of the numerical evidence demonstrates that a detonation must follow buoyancy-driven events. The GCD scenario is the most thoroughly-explored model, and the most likely result of a buoyancy-driven ignition. However, any of the variants described above will generally produce overluminous events, provided that a detonation follows the ignition of one or a small number of ignition events.

%\begin{figure}
%\begin{center}
%\includegraphics[width=1.0\columnwidth]{figures/enuc.pdf}
%\caption{A plot of the nuclear energy release $E_{\rm nuc}$ during the deflagration phase versus offset radius, for a variety of 2D and 3D simulations in the literature.  }
%\label {fig:bubble_diagram}
%\end{center}
%\end{figure}

\section {Rates, Host Galaxy Environments, and Delay-Time Distributions}
\label {sec:rates}

A long-standing problem in all major supernova channels is how to reconcile the observed SNe Ia rates with those predicted both from observations of likely stellar progenitors, as well as theoretical binary population synthesis models.  As mentioned previously, estimates of the rates of overluminous SN 1991T-like events range from 4\% - 20\% of the overall SNe Ia rate \citep {foleyetal09, lietal01}. In this section, we demonstrate that the need to account for only overluminous SNe Ia, as opposed to the entire population of SNe Ia, significantly alleviates much of the tension on single-degenerate rates. We also discuss the delay-time distributions of single-degenerate models, and compare these to the observed galactic host environments of overluminous SNe Ia.

%The SNe Ia rate is consistent with two components: a ``prompt'' component proportional to the the star-formation rate, and a ``tardy'' component proportional to the stellar mass (Scannapieco \& Bildsten (2005).

Supersoft X-ray sources have long been thought to be possible SNe Ia progenitors -- e.g. \citet {vandenheuveletal92}. However, the number of known supersoft X-ray sources \citep {distefano10} and the X-ray background \citep {gilfanovbogdan10} are discrepant with observed SNe Ia rates by 1 - 2 orders of magnitude. 

Rates derived from theoretical binary population synthesis models span a very wide range, depending on model assumptions. In particular, the derived rates are highly sensitive to the prescription of mass retention adopted, and to a lesser extent also the assumptions regarding common envelope evolution and mass transfer \citep {boursetal13, claeysetal14}. Recent work suggests that the wide range of predictions made by different BPS models employed by different groups can be connected back to varying model assumptions \citep {toonenetal14}. 

Bearing in mind these very significant uncertainties, we can compare the SNe Ia rates determined from BPS models with those observed both for normal and overluminous SNe Ia. The time-integrated BPS rates in the single-degenerate channel range from $.35  \times 10^{-4} M_{\odot}^{-1} - 1.3 \times 10^{-4} M_{\odot}^{-1}$ \citep {boursetal13}.  In contrast, the observed time-integrated SNe Ia rates are in the range of $4  \times 10^{-4} M_{\odot}^{-1} - 26  \times10^{-4} M_{\odot}^{-1}$ \citep {maozetal12, maozmannucci12,  perrettetal12, graurmaoz13}. In other words, current BPS models suggest that the single-degenerate channel accounts for only 1\% - 30\% of the total SNe Ia rate. However, this range spans the range of estimates of overluminous SNe Ia.

In contrast to buoyant ignitions, central ignitions in single-degenerate SNe Ia lead to significant amounts of deflagration energy release, and could possibly produce a range of outcomes, from failed 2002cx-like SNe Ie events through normal events. These are, however, less likely outcomes given the significantly smaller probability of ignition within the central ignition region.  We now estimate the single degenerate contribution to normal and failed SNe Ia, with two key assumptions. First, we posit that the fraction of normal and subluminous SNe Ia events stemming from the single-degenerate channel is equal to the fraction of central ignitions compared to the total. Secondly, we further assume that the overall population of overluminous events primarily originates from the single-degenerate channel. That is, if other channels, including the double-degenerate channel, do contribute to the population of overluminous SNe Ia, as we discuss below in \S \ref {sec:implications}, we assume that their contribution is negligible compared to that of single-degenerates. We place a conservative upper bound on the single-degenerate contribution by taking the fraction of central ignitions to be 5\% (including a 3 $\sigma$ variance) and the observed fraction of overluminous SNe Ia to be 20\%. Consequently, we infer the contribution of the single degenerate channel to both normal SNe Ia and failed 2002cx events to be no greater than 1\% of the overall SNe Ia rate. The low probability of central ignition generally implies the contribution of single-degenerates to normal and failed SNe Ia events must be small. In particular, if other channels also contribute significantly to the overluminous SNe Ia rate, the single-degenerate contribution to normal and failed SNe Ia events will be even smaller than our conservative upper-bound. We thus suggest that the progenitors of normal and failed 2002cx events must  primarily originate from other channels -- most likely the double-degenerate and helium donor channels.

Many studies have confirmed that overluminous SNe Ia events arise primarily in late-type galaxies \cite {howell01}, and that late-type galaxy SNe Ia are overluminous in comparison  to those which arise early-type galaxies \citep {hamuyetal00,  sullivanetal06, lampeitletal10, guptaetal11, johanssonetal13, childressetal13}. The late-type galaxy host environment of overluminous SNe Ia can be explained in the context of the single-degenerate channel by noting that the single-degenerate DTD is relatively prompt in comparison to the double-degenerate channel, with delay times starting at 40 - 100 Myr, depending on model assumptions \citep {nomotoetal07, ruiteretal09, toonenetal14}.  Moreover, the single-degenerate channel DTD is distinguished from the double-degenerate channel in the need to have a main-sequence or red giant donor; consequently the single-degenerate DTD exhibits a sharp cutoff at 2 - 3 Gyr, independent of model assumptions. Hence, single-degenerate SNe Ia are naturally expected to be present in relatively young stellar populations, and absent from very old ($>$ 2 - 3 Gyr) stellar populations, as has long been noted -- see e.g. \cite {dellavalle94}. However, we also note that massive C/O primaries are naturally more luminous in both the double-degenerate and double-detonation models as well. Such massive C/O WD primaries arise from more massive intermediate-mass stellar progenitors, with prompt delay times ($\lesssim 100$ Myr) \citep {ruiteretal13}, and therefore also naturally arise in late-type galaxies. Hence the correlation between peak brightness and host galaxy environment is one which is, broadly speaking, expected for single-degenerate SNe Ia as well as double-degenerate and double-detonation SNe Ia. Consequently, in order to better understand the correlation between the host environments and luminosities of SNe Ia, additional observational constraints which directly test the single-degenerate channel are needed, as we discuss in the discussion in \S \ref {sec:implications}.

In summary, if the single-degenerate channel must account only for overluminous SNe Ia, and not the population of normal SNe Ia as a whole, the estimated rates derived both from the known population of supersoft X-ray sources and from theoretical BPS models are much more closely aligned. While much work remains to be done, particularly with regards to improving the mass retention efficiency physics of the BPS models, the overall consistency between the rates of the single-degenerate sources and overluminous SNe Ia rates motivates refined observations to test this connection further.

\section {Discussion}
\label {sec:implications}

\citet {simetal13} have demonstrated that synthetic spectra of single and few-bubble offset ignition simulations  are nominally good matches to SN 1991T-like SNe Ia events, according to the standard criterion of a sufficiently high spectral cross-correlation coefficient \citep {blondintonry07}. The synthetic spectra support the view that single-degenerate supernovae are promising candidates for SN 1991T-like events. Because the deflagration stage is most particularly sensitive to the central density of the progenitor, further exploring this model sensitivity may help clarify the quality of the cross-correlation of models with observations, and provide theoretical spectral templates to compare against observations.

 In principle, the presence of a main-sequence or red-giant companion star is detectable in early-time light curve distortions arising from the shocked gas of the companion under favorable viewing angles \citep {kasen10}. Extensive light curve data from the Sloan Digital Sky Survey and  the Supernova Legacy Survey \citep {haydenetal10, biancoetal11} failed to produce evidence for such distortions, and constrained the fraction of red giant companions to be 10\% - 20\% of the overall population of SNe Ia -- within the range expected for overluminous SNe Ia. Further exploration of these early light curve signatures focusing upon overluminous SNe Ia events may be an interesting avenue for future work. However, the recent identification of the diversity of the distribution of $^{56}$Ni in the outermost ejecta of SNe Ia, reflected in a range of early light curve profiles \citep {firthetal15} will make it more challenging to characterize the possible presence of shocked gas from the companions in the data. 

The presence of any significant amount of hydrogen in the nebular phase of a SNe Ia spectrum would be a ``smoking gun'' piece of evidence of stripped hydrogen from the companion star in favor of the single-degenerate channel, but has so far eluded detection. Tight observational constraints have been placed on the hydrogen abundance in SNe, as determined from the absence of any observable H$\alpha$ emission lines in the nebular phase spectra of normal Ia.  \citet {leonard07} constrained the amount of H in the normal SN Ia 2005cf and the slightly subluminous SN Ia 2005am to be less than $10^{-2} M_{\odot}$.  \citet {shappeeetal13a}, \citet {lundqvistetal15}, and \citet {grahametal15} applied similar methods to place the tightest-known constraints on the amount of H in SN 2011fe, at $< 3 \times 10^{-4} M_{\odot}$.  Although theoretical caveats remain to be investigated that may serve to obscure H$\alpha$, even at late times -- see e.g., \citet {leonard07} -- current observational limits on SN 2011fe and other normally-bright SNe Ia are well beneath the range predicted for nearly all types of red giant and main sequence, including helium main sequence companions \citep {mariettaetal00, pakmoretal08, panetal13, panetal14}, with the possible exception of M-dwarfs \citep {wheeler12}. Crucially, however, we note that the most important constraints on H$\alpha$ in the nebular phase of SNe Ia are all derived from normal SNe Ia. The presence of H$\alpha$ in overluminous SNe Ia remain unconstrained by data of comparable quality. 

Consequently, one of the most important tests of the contribution of single-degenerate supernovae to overluminous SNe Ia will be a deep probe for H$\alpha$ in the nebular optical spectra of low-redshift overluminous SNe Ia. Additional recent work suggests P$_\beta$ in the near-infrared post-maximum \citep {maedaetal14} may also be a possible strong discriminant of the companion. Ongoing surveys including ASASSN will identify a number of low-redshift SN 1991T-like SNe Ia which will enable a comprehensive survey in a number of events. Already several good candidates, including  ASASSN-14lw, PESSTO ESO 154-G10, and ASASSN-15bc at $z < .04$ have recently been discovered, and will be prime candidates for nebular-phase follow-up in the near future. 

The relatively rare class of SNe Ia-CSM demonstrates evidence for H$\alpha$ emission. SNe Ia-CSM originate exclusively in late-type galaxies and irregulars and are overluminous, with $-21.3 \leq M_R \leq -19$ peak mag \citep {silvermanetal13}. Moreover, recent evidence points to an association of SNe Ia-CSM and overluminous SNe 1991T-like events \citep {leloudasetal15}. In contrast, multi-wavelength observations, including upper-bounds of radio and X-ray emission from the normal SN Ia 2011fe, place tight constraints on the possible stellar companions and the circumstellar environment around normal SNe Ia \citep {lietal11, brownetal12, horeshetal12, bloometal12, marguttietal12, chomiuketal12}. Similar observations, carried out for an overluminous SNe Ia event at a comparable distance, are the most promising site of the first multi-wavelength signature from a SN Ia.

% Livio 1994
% http://adsabs.harvard.edu/abs/1994ApJ...423L..31D

Another strong piece of evidence in favor of the single-degenerate channel would be the detection of the companion star in a Type Ia supernova remnant. However,  until very recently, there has been a general lack of supporting evidence of these so-called ``ex-companion'' stars. In particular, observers have found an absence of ex-companions in SNRs SN2006dd and SN2006mr in NGC 1316  \citep {maozmannuccietal08}, SN 1006 \citep {gonzalezhernandez12, kerzendorfetal12}, SNR 0509-67.5 \citep {schaeferpagnotta12}, SNR 0519-69.0 \citep {edwardsetal12}, and SN 1604 Kepler \citep {kerzendorf_etal_2014}. A candidate ex-companion in SN 1572 Tycho has been identified \citep {ruizlapuente04}, but has been questioned as our knowledge of the properties of ex-companions improved \citep {shappeeetal13b, panetal14}.  Very recent work has revealed a number of possible ex-companions in the SNe Ia remnants SNR 0505-67.9 and SNR 0509-68.7 \citep {pagnottaschaefer15}. While the sample size of investigated remnants is so far very small, if one or both of the ex-companions in the remnants 0505-67.9 and SNR 0509-68.7 are confirmed, the frequency of ex-companions in known remnants will be consistent with the frequency of overluminous SNe Ia compared to SNe Ia overall. Further work will help clarify these events and demonstrate if in fact this is the case. 

Additionally, we note that light echoes have identified a small handful of galactic remnants as normal SNe Ia  -- see e..g, \citet {krauseetal08}.  SNR 0509-67.5 is the only extragalactic remnant which has a subtype identification confirmed by light echoes; it has been identified as an overluminous supernova remnant \citep {restetal08, badenesetal08}. The overluminous subtype classification of  SNR 0509-67.5 presents an interesting challenge to the framework presented in this paper. This event appears to be near the boundary of normal and overluminous events; of the six best spectral template matches to  SNR 0509-67.5, three are SN 1991T-like events, and three are normal, though the normal events are not as in good agreement as the overluminous events.  Alternatively, it is also possible that double-degenerates may give rise to overluminous SNe Ia alongside single-degenerates. Spectral classification of additional remnants, most crucially 0505-67.9 and SNR 0509-68.7, if confirmed to harbor ex-companions,  as well as 3C 397 (see below) will yield crucial data confronting theory.

%% RTF Added on resubmission, 2/25/15.

Lastly, one of the most direct diagnostics of the amount of $^{56}$Ni synthesized in a SNe Ia event is the hard X-ray Fe K-$\alpha$ line in young remnants. The hard X-ray portion of the spectrum also contains adjacent K-$\alpha$ lines of stable iron-peak elements including $^{58}$Ni, $^{55}$Mn, and $^{52}$Cr, which are produced by electron captures in significant abundances only at the high densities of Chandrasekhar-mass white dwarfs \citep {seitenzahletal13}, and are a characteristic nucleosynthetic signature of the single-degenerate channel.  Direct detection of the 5.9 keV Mn K-shell line produced by the decay of radioactive $^{55}$Fe may also be possible in the first several years following nearby SN Ia events\citep {seitenzahletal15}.  After the initial submission of this paper, significant levels of stable Fe-peak elements in Suzaku X-ray data were announced  for SNR 3C 397 \citep {yamaguchietal15}, yielding the strongest supporting evidence to date of a single-degenerate SN Ia. Furthermore, 3C 397 has one of the highest measured X-ray luminosities in the Fe K-$\alpha$ line of all SNe Ia remnants, suggestive of an overluminous nucleosynthetic yield of $\sim 1.0 M_{\odot}$ of $^{56}$Ni \citep {yamaguchietal14, patnaudeetal15}, though more work is needed to compare these recent observations against theoretical models and more precisely determine the likely $^{56}$Ni yield of this event.  A crucial test of the self-consistency of the single-degenerate paradigm will be to determine if 3C 397 harbors an ex-companion. A light echo spectra for 3C 397 of sufficiently high quality to determine the subtype of the event will be a further independent confirmation of the luminosity. Theoretical work confronting three-dimensional SNe Ia explosion models, evolved through the young SNR remnant phase, with this recent hard X-ray spectral data will further shed light on the single-degenerate explosion mechanism. Additionally, and most crucially, further observations and modeling are required to detect stable Fe-peak elements in other remnants and to quantify their  $^{56}$Ni  yields, which will provide crucial constraints on the relative contributions of both the single-degenerate and double-degenerate channels to the overall SNe Ia rate, and inform optical light curve models.
  
Combining all available observational and theoretical evidence, the picture that emerges is that Chandrasekhar-mass white dwarfs, long thought to be the origin of the standardizable candle, may instead be the anomalies.  In this view, the stellar progenitors of single-degenerate SNe Ia are primarily the rapidly-accreting supersoft X-ray sources, leading to overluminous SN 1991T-like events, with recurrent novae responsible for the smaller class of SNe Ia-CSM events.  Normal SNe Ia events are very likely to instead originate primarily from a combination of the double-degenerate channel and double-detonations. Consequently, the absence of evidence for the single-degenerate channel, derived almost entirely from the more commonplace normal SNe Ia, is not altogether surprising. All evidence points to the conclusion that single-degenerate SNe must be rare in comparison to the total SNe Ia rate, the proverbial needle in the haystack. In order to either confirm or strongly rule out the single-degenerate channel, one must know where in the haystack to look. The class of overluminous SNe Ia represents the most likely outcome of the single-degenerate channel, and future observations focused upon overluminous events may help elucidate their nature.

{\bf Acknowledgements}  The authors would like to thank the anonymous reviewer for numerous helpful suggestions which have strengthened the paper considerably. RTF thanks Pranav Dave, Rahul Kashyap, Douglas Leonard, Giorgos Leloudas, Jeffrey Silverman, Martin van Kerkwijk, Hirosha Yamaguchi, and Carles Badenes for fruitful conversations. RTF also thanks Avi Loeb, James Guillochon, and the Harvard-Smithsonian Center for Astrophysics Institute for Theory and Computation for a kind invitation to present an early version of this work. K.J. acknowledges support from NSF Grant DMS-0802974.

\bibliography{converted_to_latex}

\appendix

In this appendix, we fill in the physical and mathematical modeling of the birth of a flame bubble during the simmering phase of Chandrasekhar-mass white dwarfs. We assume that the flame bubble begins with ignition of a spherical bubble with radius $R_0$. The assumption of sphericity is an excellent one for small flame bubbles less than the fire-polishing scale $\lambda_{\rm fp}$ defined in the text, since the action of the flame will rapidly erase any non-spherical distortions. In this case, the bubble radius $R (t)$ satisfies the equation :

\begin {equation}
\dot {R} = \left ({1 + A \over 1 - A }\right) S_l
\label {rdot}
\end {equation}
where the Atwood number $A$ is defined in the text. Numerical simulations demonstrate that the distribution of offset radii occurs well within the central pressure scale height ($\sim 400$ km) of the white dwarf, so the density stratification within the ignition region may be neglected to a good degree of approximation. With the zero time chosen as the ignition of the bubble, the bubble radius $R$ is then simply expressed as 

\begin {equation}
R (t) = R_0 + \left (1 + A \over 1 - A \right) S_l t
\end {equation}
where $S_l$ is the laminar flame speed.

The difference between the density of the ash in the bubble and the density of the surrounding fuel will soon cause the bubble to buoyantly rise.  Additionally, the bubble is subject to the added mass effect, in which the action of displacing the surrounding fuel causes upward inertia  in the bubble, effectively increasing its buoyancy \citep {landaulifshitz59}. For a spherical bubble, the added mass is precisely $1/2$ the mass of the fuel displaced.
%Cut mention of quadratic drag here -- neglected in comparison to linear frag 
%Moreover, the bubble's motion through the surrounding fuel will also cause a drag force to be exerted on it, slowing its upward rise.  
We can summarize these effects with the Morison Equation \citep {Brennen} :

\begin {equation}
%\begin {align*}
{d \over dt} \left [ {4\over 3} \pi R^3 \left ({\rho_a + {1 \over 2}  \rho_f} \right) v \right] = 
{4 \over 3} \pi R^3  \left( \rho_f - \rho_a \right) g 
\label {morisoneqn}
%\end {align*}
\end {equation}
here $t$ is time, $R$ is the radius of the bubble, $\rho_a$ is the ash density of the bubble, $\rho_f$ is the fuel density of the bubble, $v$ is its rise speed, $g$ is local gravitational acceleration.

The left-hand side of equation \ref {morisoneqn} is the time derivative of the inertia.  The first term is the inertia of the bubble itself, whereas the second is the inertia contributed by the added mass effect, exactly equal to half that of the fuel density displaced by the spherical bubble.  In contrast, the right hand side gives the buoyancy force.  Taking the time derivative of inertia yields both an acceleration and a drag term, which arises from the growing mass of the bubble. One can easily demonstrate that the net acceleration of the bubble is :

\begin {equation}
{d v\over dt} = a_b - a_d
\end {equation}
where the acceleration due to buoyancy, $a_b$, and drag, $a_d$, are :

\begin {equation}
a_b = 2 K_a g {A \over 1 - A}
\end {equation}
\begin {equation}
%a_d = {1 + A \over 1 - A}\left[ {3 \over 8} K_a C_D {v^2 \over R}  + 3 S_f {v \over R} \right]
a_d = 3 \left( {1 + A \over 1 - A} \right)   {S_l v \over R} 
\label {drageqn}
\end {equation}
and $K_a$ is a factor taking into account the added-mass effect, and is defined as

\begin {equation}
K_a = {1 \over {1 + {1 \over 2} {{1 + A} \over 1 - A} } } 
\end {equation}

We note that our equations \ref {rdot} and \ref {morisoneqn} are nearly identical to those given by \citet {aspdenetal11}, but differ in one respect. Our model takes the flame bubble growth at the laminar flame speed, which is appropriate to the early-growth of the flame bubble. In contrast, \citet {aspdenetal11} also include the effect of entrainment, which adds a term proportional to the bubble rise speed $v$ to equation \ref {rdot}. Additionally, \citet {aspdenetal11} use two free parameters to fit the effective buoyancy and drag of their model to numerical simulations, whereas our approach has no free parameters. Nonetheless, the two approaches are in good agreement during the initial stages of bubble growth following ignition, where the bubble rise speed is less than the laminar flame speed.

%From Equations 5 and 6, we can integrate to find the rising speed of the bubble.

%%
%\begin {equation}
%v = v_o + \int_0^t (a_b - a_d) dt
%\end {equation}
%
%$v_o$ is the initial rise velocity of the bubble.

We now derive an equation of motion for the bubble offset radius $r (t)$.  Buoyant acceleration is independent of the size of the bubble, whereas the drag acceleration depends inversely on bubble radius. Consequently, the motion of a small flame bubble is strongly influenced by drag. 
%Further, the second, linear drag term in equation \ref {drageqn} dominates the first, linear drag term during the initial growth of the bubble, when the buoyant velocity $v < S_l$. 
The ignition points obtained within numerical simulations are typically very small (of order km), and are themselves likely limited by numerical resolution. Hence, we are motivated to taking the limit of zero initial flame bubble radius, going over to the self-similar limit, with the bubble radius simply becoming $R (t) = [(1 + A) / (1 - A)] S_l t$. 

%\begin {equation}
%v = {2 \over 3} K_a \left ({A \over 1 + A}   \right) g t
%\end {equation}

Writing the gravitational acceleration in the central, nearly-uniform density portion of the WD as $g = g_0 r$, the offset radius of a self-similar bubble satisfies the following equation :

\begin {equation}
\left [ {d^2 r\over dt^2} -  2 K_a \left (A \over 1 - A   \right) g_0 r\right] t + 3   {dr \over dt}  = 0
\end {equation}

The equation is singular at $t = 0$ due to the assumption of zero bubble radius at ignition, but is well-behaved provided the bubble begins at rest with respect to the background flow -- a reasonable assumption which we adopt.  The solution to this equation may be found by a power-series expansion in time. Under the assumption of zero initial velocity, it is easily demonstrated that the odd terms in this series all identically vanish. To leading order,

\begin {equation}
r (t) = r_0 + {1 \over 2}  \left({ A\over 3 - A}  \right) g_0 r_0 t^2
\end {equation}

This implies an effective acceleration equal to 

\begin {equation}
g_{\rm eff} =  \left({A\over 3 - A}  \right) g_0 r_0
\end {equation}
which is the value we use in the text.

\end{document}